\documentstyle[12pt]{article}
\newcommand{\be}{\begin{eqnarray}}
\newcommand{\ee}{\end{eqnarray}}
\newcommand{\D}{{\cal D}}
\newcommand{\dslash}{\partial \hskip -0.5em /}
\newcommand{\Dslash}{D \hskip -0.7em /}
\newcommand{\Vslash}{V \hskip -0.7em /}

\begin{document}
\rightline{UNIT\"U--THEP-19/1992}
\vskip 2truecm

\centerline{\large\bf The chiral anomaly and electromagnetic pion decay}
\centerline{\large\bf in effective quark models}
\vskip 2cm
\centerline{R.\ Alkofer and H.\ Reinhardt}
\centerline{Institute for Theoretical Physics}
\centerline{University of T\"ubingen}
\centerline{Auf der Morgenstelle 14}
\centerline{D-W7400 T\"ubingen, Germany}
\vskip 1cm
\vskip 6cm

\centerline{\bf ABSTRACT}
\medskip
Within renormalizable theories the triangle
diagram correctly reproduces the lifetime of the neutral pion $\pi ^0$.
However, effective relativistic quark models which contain a second scale
besides the quark mass seem unable to give the correct value for this
quantity. In the Instanton
Liquid Model this second scale is the mean instanton radius, in the
NJL model it is the cutoff. The role of the cutoff within a path
integral derivation of the chiral anomaly is clarified and the
importance of higher order terms contributing to the $\pi ^0$ decay width
and other anomalous processes is pointed out.

\vfill
\eject

Effective chiral quark Lagrangians have been very successful in modeling hadron
physics. In particular, Nambu--Jona-Lasinio (NJL) like models [1] are
able to reproduce meson spectra, decays and even scattering lengths with
a few input parameters adjusted [2,3,4,5,6].
However, there is one case where these models seem to fail:
the decay $\pi ^0\to 2 \gamma $ [6].
It was as early as 1949 Steinberger calculated the correct
lifetime of the neutral pion from the triangle diagram Fig.~1 [7]. He
interpreted the internal fermion lines as nucleons whereas nowadays we
identify them with quarks. It is well-known that the triangle diagram is
tightly related to the famous Adler-Bell-Jackiw (ABJ) [8] or chiral anomaly.

\begin{figure}[b]
\begin{picture}(400,120)
\thicklines
\put(80,60){$\pi^0$}
\multiput(100,60)(10,0){6}{\line(1,0){5}}
\put(200,100){\vector(-1,-1){20}}
\put(160,60){\line(1,1){20}}
\put(160,60){\vector(1,-1){20}}
\put(180,40){\line(1,-1){20}}
\put(200,20){\vector(0,1){40}}
\put(200,60){\line(0,1){40}}
\multiput(202.5,20)(10,0){6}{\oval(5,3)[t]}
\multiput(207.5,20)(10,0){6}{\oval(5,3)[b]}
\multiput(202.5,100)(10,0){6}{\oval(5,3)[t]}
\multiput(207.5,100)(10,0){6}{\oval(5,3)[b]}
\put(280,20){$\gamma $}
\put(280,100){$\gamma $}
\end{picture}
\caption{The triangle diagram: only contribution to $\pi ^0 \to 2\gamma
$ in renormalizable theories and leading order contribution in effective
models.}
\end{figure}
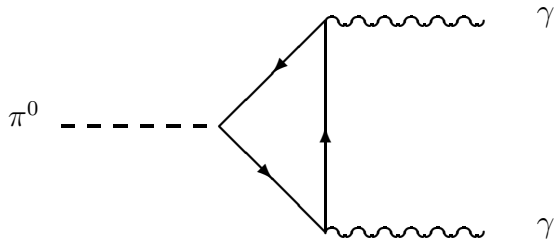

Common to all type of effective relativistic quark models is the
spontaneous breaking of chiral symmetry which is reflected in dynamical
generation of constituent quark masses even for vanishing current
masses. In the chiral limit these constituent quark masses set the scale.
In this letter we shall demonstrate that the failure of effective quark
models to give the
correct value of the $\pi^0$ lifetime from the triangle diagram is
connected to the appearance of a second scale, which is given by the
inverse of the mean instanton radius $\rho $ in the so-called Instanton Liquid
Model [9] and by the cutoff $\Lambda $ in the NJL model.
In both models the triangle diagram (Fig.~1)
gives the correct leading-order result in a $1/N_c$ expansion $N_c$ being
the number of colours [10].
However, the numerical value for the decay amplitude is only approximately two
thirds of the experimental value.
This can be quite easily seen. In the
chiral limit ($m_\pi =0$)
and for on--shell photons ($q^2=0$) the Feynman diagram of Fig.\ 1
leads to the integral (in Euclidian space)
\be
I=4 N_c \int \frac {d^4k}{(2\pi )^4} \frac {M^2(k)}{(k^2+M^2(k))^3} .
\ee
This integral has quite interesting properties. First, it is finite.
Second, assuming the quark constituent mass $M$ to be momentum--independent
and the domain of integration to be unrestricted one obtains
$I=N_c/(8\pi ^2)$ independent of $M$. This value of $I$ yields then the
correct $\pi^0$ lifetime. In the Instanton Liquid Model the constituent
mass depends on the momentum [10]
\be
M(k)&=& M(0) f(k\rho /2)
\nonumber \\
f(z)&=&\frac d {dz} (I_0(z)K_0(z)-I_1(z)K_1(z))
\ee
where $I_{0,1}$ and $K_{0,1}$ are modified Bessel functions. Eq.\ (2)
simply states that $M$ is exponentially decreasing for momenta $k\gg
1/\rho $, {\it i.e.} for momenta much larger than the inverse of the
mean instanton radius $\rho $ the integrand of $I$ is strongly damped.
Using now the
standard parameters of the Instanton Liquid Model, instanton density
$N/V= 1 {\rm fm}^{-4} = (200 {\rm MeV})^4$ and mean instanton radius
$\rho = \frac 1 3 {\rm fm} = 1/(600 {\rm MeV})$, one obtains only 63 per
cent of the value which one has for a constant constituent mass. Note
that the pion life time then comes out too large by more than a factor of two.
This is because the integral $I$ enters quadratically in the expression
for the $\pi ^0$ decay width. It is also obvious that $I$ will be too
small in NJL like models where $M$ is a constant, however, the
integral $I$ is cutoff by some regularization procedure. Numerical
values may be taken from ref.\ 6 where for a Euclidian sharp cutoff
60\% of the experimental value
for the $\pi ^0$ decay width are found for a reasonable choice of
parameters. Note also that this number depends on the regularization
method. Using proper time regularization [11] we estimated the
corresponding value to be 75 \%.

In order to illuminate the origin of
this discrepancy it is convenient to employ
Fujikawa's path integral derivation of the chiral anomaly [12]. We shall
do that for the NJL model using proper time regularization.
Under a local chiral transformation
\be
\tilde q & = & e^{i\Theta (x) \gamma_5} q
\ee
the measure of a fermion path integral is in general not invariant
\be
\D \tilde q \D \bar {\tilde q} &=& J(\Theta ) \D q \D \bar q .
\ee
This is in particular true in the presence of a vector field which in the
present case is the photon field. The Jacobian $J(\Theta )$ can be expressed
as [12]
\be
J(\Theta ) &=& \exp \left( -2i \int d^4x \Theta (x) B(x) \right)
\ee
where $B(x)$ is, in a na\"\i ve sense, the trace of
$\gamma_5$\footnote{The trace includes Dirac, colour and flavour indices.}
\be
B(x) = {\rm tr } \gamma _5 \delta _{reg} (x,x) .
\ee
This quantity can be only calculated using a regularization procedure.
Adopting the proper time method we define
\be
B_\Lambda (x) = {\rm tr} \left( \gamma_5 \langle x |
\int _{1/\Lambda^2}^{\infty}
d\tau \Dslash ^\dagger \Dslash \exp (-\tau \Dslash ^\dagger \Dslash)
|x \rangle \right)
\ee
where the Dirac operator
\be
i\Dslash = i (\dslash + \Vslash ) - M
\ee
describes a quark with constituent mass $M$ moving in the external
photon field $V_\mu$. Note that the parameter
integral in eq.\ (7)  may be written as incomplete gamma function
$\Gamma (1,\Dslash ^\dagger \Dslash /\Lambda^2)$. For $\Lambda^2 \to
\infty $ it becomes simply $\Gamma (1,0) = \Gamma (1) =1$. On the other
hand, even in a renormalizable theory the limit $\Lambda^2 \to \infty $
may not be taken immediately because the right hand side of eq.\ (6)
is an ill-defined
object. Note also that the proper time method is a gauge invariant
regularization procedure.

Using the formula
\be
\int _{1/\Lambda^2}^{\infty} d\tau e^{-\tau A} &=&
\int _{1/\Lambda^2}^{\infty} d\tau e^{-\tau (A_0+\Delta A)}
\nonumber \\*
&=&\int _{1/\Lambda^2}^{\infty} d\tau e^{-\tau A_0}
-\int _{1/\Lambda^2}^{\infty} d\tau \tau \int _0^1 d \zeta e^{ -\tau
\zeta A_0}\Delta A e^{ -\tau (1-\zeta ) A_0}
\nonumber \\*
&+& \int _{1/\Lambda^2}^{\infty} d\tau \tau ^2 \int _0^1 d \zeta \int
_0^{1-\zeta } d \eta e^{ -\tau \eta A_0} \Delta A e^{ -\tau
(1-\zeta -\eta )A_0}\Delta A e^{ -\tau \zeta A_0} + \ldots
\ee
and
\be
\Dslash ^\dagger \Dslash = \partial^\mu \partial _\mu +M^2
+\{\partial^\mu, V_\mu \} +\frac 1 4 [ \gamma _\mu , \gamma _\nu ]
F^{\mu \nu} +V^\mu V_\mu
\ee
we expand $B_\Lambda $ up to second order in the photon field $V$. Due
to the abelian nature of $V$ the field strength tensor $F^{\mu \nu}$ is
linear in $V$. There are two terms with non--vanishing
Dirac trace. Both are proportional to
\be
{\rm tr}_D (\gamma _5 \frac 1 4 [ \gamma ^\mu ,\gamma ^\nu ]
\frac 1 4 [ \gamma ^\kappa , \gamma ^\lambda ] )
F _{\mu \nu} F _{\kappa \lambda }
= \epsilon ^{\mu \nu \kappa \lambda }  F _{\mu \nu} F _{\kappa \lambda }
= 2 \tilde F ^{\mu \nu} F _{\mu \nu }
\ee
where we have used the definition of the dual field tensor $\tilde F
^{\mu \nu}$. After some straightforward algebra we obtain
\be
B_\Lambda (x)= 2N_c \tilde F ^{\mu \nu} (x)F _{\mu \nu } (x)
\langle x | \Bigl(
\int _{1/\Lambda^2}^{\infty} d\tau \tau ^2 A_0 \int _0^1 d \zeta \int
_0^{1-\zeta } d \eta e^{ -\tau \eta A_0} e^{ -\tau (1-\zeta -\eta )A_0}
e^{ -\tau \zeta A_0}
\nonumber\\*
-\int _{1/\Lambda^2}^{\infty} d\tau \tau \int _0^1 d \zeta ^{ -\tau
\zeta A_0}e^{ -\tau (1-\zeta )A_0} \Bigr) |x\rangle + {\cal O}(V^4)
\ee
where $A_0=\partial^\mu \partial_\mu +M^2$. The matrix element in eq.\ (12)
is most easily calculated in momentum space
\be
B_\Lambda (x) = 2N_c \tilde F ^{\mu \nu} (x) F _{\mu \nu }(x)
\int \frac {d^4k}{(2\pi )^4} \int _{1/\Lambda^2}^{\infty} d\tau
\left( \frac 1 2 \tau ^2 (k^2+M^2) -\tau \right) e^{-\tau (k^2+M^2)}
+ {\cal O}(V^4)
\nonumber \\*
\ee
where we have used $\int _0^1 d \zeta \int _0^{1-\zeta } \eta = 1/2$.
Evaluating the momentum integral one obtains
\be
B_\Lambda &=& \frac {N_c}{16\pi ^2} \tilde F ^{\mu \nu} F _{\mu \nu }
M^2 \int _{1/\Lambda^2}^{\infty} d\tau e^{-\tau M^2} + {\cal O}(V^4)
\nonumber \\*
&=& \frac {N_c}{16\pi ^2} \tilde F ^{\mu \nu} F _{\mu \nu }
e^{-\frac {M^2}{\Lambda ^2} } + {\cal O}(V^4).
\ee
It may be easily checked by simple power counting that the ${\cal
O}(V^4)$ and higher order contributions vanish as $\Lambda \to \infty $.
Therefore in a renormalizable theory we obtain the well--known result
\be
B(x) = \frac {N_c}{16\pi ^2} \tilde F _{\mu \nu } F ^{\mu \nu }
\ee
which yields for the Jacobian $J(\Theta )$
\be
J(\Theta ) = \exp \left( -i \frac {N_c}{8\pi ^2} \int
d^4 x \Theta (x) \tilde F_{\mu \nu } F^{\mu \nu }  \right) .
\ee
 From here several well-known relations as {\it e.g.} the
anomalous Ward identity ($m^0$ being the current quark mass and $e$ the
electric charge)
\be
\partial _\mu j_5^\mu = 2im^0 j_5 -i\frac {N_c}{8\pi ^2} e^2
\tilde F_{\mu \nu } F ^{\mu \nu }
\ee
can be derived using standard techniques.

The crucial observation is that eq.\ (15) is {\bf not valid} if the cutoff
$\Lambda $ is kept finite.
In this case
one has to evaluate $B_\Lambda $ to all orders in the photon field
$V_\mu $. For the decay amplitude this means that instead of
\be
\epsilon ^\mu _1 \epsilon ^\nu _2 T_{\mu \nu } (q^2) =
\frac {m_\pi ^2 -q^2}{m_\pi ^2f_\pi ^2} e^2
\langle \gamma (k_1,\epsilon _1)\gamma (k_2,\epsilon _2) \vert
\frac {N_c}{16\pi ^2} \tilde F _{\mu \nu } F ^{\mu \nu } \vert
0 \rangle
\ee
one has to calculate
\be
\frac {m_\pi ^2 -q^2}{m_\pi ^2f_\pi ^2} e^2
\langle \gamma (k_1,\epsilon _1)\gamma (k_2,\epsilon _2) \vert
B_\Lambda (x) \vert 0 \rangle
\ee
using the expression (7) with finite cutoff.
In principle, all orders in $V$ have to be taken into account.
However, for $\Lambda \gg M$ a
few additional terms should be sufficient in order to yield most of the
missing part of the decay amplitude.

In other effective models where instead of the cutoff one has a
momentum--dependent constituent quark mass the momentum integral in
eq.\ (13) is more complicated. As the quark mass is decreasing with
increasing momenta one will always obtain attenuation of the leading
order term. On the other hand, power counting for the ${\cal O} (V^4)$
does only provide the fact that these contributions are suppressed by
inverse powers of the typical momentum scale inherent to
the constituent quark mass.

Let us summarize: If one works with bare quarks, {\it i.e.}
momentum--independent quark masses, and calculates the ABJ
anomaly within a renormalizable theory in which the cutoff will be
removed at the end of the calculation, $\Lambda \to
\infty $, only the triangle diagram Fig.~1 contributes and reproduces the
correct $\pi ^0$ life time. However, if one works in an effective theory
with dressed quarks, {\it i.e.} with finite constituent quark
mass and finite cutoff (or some other scale which makes the constituent
quark mass decrease at high momenta), not only the triangle
contributes. In principle, the chiral anomaly $B_\Lambda $ has to be
calculated exactly. These additional contributions are important for
$\pi ^0 \to 2 \gamma $ and all other processes involving the anomaly if
one uses effective models. Including the
next-to-leading order of the photon field
should significantly improve the results for the $\pi^0$ life time
in these effective models thereby resolving the problems pointed out
in ref.\ 6.

\vskip 1cm
\leftline{\it Acknowledgements}
We thank C.\ Weiss for helpful discussions and a critical reading of the
manuscript.

\vfill\eject
\leftline{\it References}
\begin{enumerate}
\item
Y.\ Nambu and G.\ Jona-Lasinio, Phys.\ Rev.\ {\bf 122} (1961) 345;
{\bf 124} (1961) 246.
\item
D.\ Ebert and H.\ Reinhardt, Nucl.\ Phys.\ {\bf B271} (1986) 188.

\item
H.\ Reinhardt and R.\ Alkofer,  Phys.\ Lett.\ {\bf 207B} (1988) 482;

R.\ Alkofer and H.\ Reinhardt, Z.\ Phys.\ {\bf C45} (1989) 245.

\item
S.\ Klimt, M.\ Lutz, U.\ Vogl and W.\ Weise, Nucl.\ Phys.\ {\bf A516}
(1990) 429;
U.\ Vogl, M.\ Lutz, S.\ Klimt and W.\ Weise, Nucl.\ Phys.\ {\bf A516}
(1990) 469.

\item
V.\ Bernard, U.-G.\ Mei\ss ner, A.\ H.\ Blin and B.\ Hiller,
Phys.\ Lett.\ {\bf 253B} (1991) 443.
\item
A.\ Blin, B.\ Hiller and M.\ Schaden, Z.\ Phys.\  {\bf A331} (1988) 75.


\item
J.\ Steinberger, Phys.\ Rev.\ {\bf 76} (1949) 1180.

\item
S.\ Adler, Phys.\ Rev.\ {\bf 177} (1969) 2426;

J.\ S.\ Bell and R.\ Jackiw, Nuov.\ Cim.\ {\bf 60A} (1969) 47.

\item
see {\it e.g.}: E.\ V.\ Shuryak, Phys.\ Rep.\ {\bf 115} (1985) 152.

\item
D.\ I.\ Diakonov and V.\ Y.\ Petrov, Nucl.\ Phys.\ {\bf B272} (1986)
457.

\item
J.\ Schwinger, Phys.\ Rev.\ {\bf 82} (1951) 664.

\item
K.\ Fujikawa, Phys.\ Rev.\ {\bf D21} (1980) 2848.

\end{enumerate}

\end{document}